\documentclass[12pt]{article}
\usepackage[english]{babel}
\usepackage{graphicx}
\usepackage{amsmath}
\usepackage{amsfonts}
\usepackage{amssymb}

\begin{document}

\title{A new Physics to support the Copernican system. Gleanings 
from Galileo's works}

\author{Giulio Peruzzi\thanks{Department of Physics, University of Padua, e-mail: peruzzi@pd.infn.it.}}

\date{}

\maketitle

\begin{abstract}
Galileo's support to the Copernican theory was 
decisive for the revolutionary astronomical discoveries he achieved 
in 1610. We trace the origins of Galileo's conversion to the 
Copernican theory, discussing in particular the ``Dialogo de 
Cecco di Ronchitti da Bruzene in perpuosito de La Stella Nuova''. 
Later developments of Galileo's works are briefly treated.
\end{abstract}

The use of the telescope alone doesn't explain the revolutionary 
astronomical discoveries achieved by Galileo from the end of 
1609 onwards. To look doesn't mean to see, and the ``sensate esperienze'' 
must integrate observation and experimentation. Galileo looks 
and sees because in preceding years he had freed himself from 
prevailing convictions and he had progressively become aware that 
the facts he was studying both in the Heavens and on the Earth 
went in the direction of confirming the Copernican system. 

It is well known that one of the first evidences of his adherence 
to Copernicanism lies in a letter to Kepler written on 4$^{th}$ August 
1597.\footnote{\textit{Le Opere di Galileo Galilei. Edizione Nazionale}, 
vol. X, pp. 67-8, p. 68. Hereinafter referred to simply \textit{Opere}.} 
Galileo is however well aware that the Copernican system, unlike 
the Aristotelian-Ptolemaic system, lacks a physics of its own. 
It is not by chance that in the years preceding the use of the 
telescope, his researches were devoted to both astronomy 
and the study of local motions. In this sense, it is emblematic 
that Galileo analyses the \textit{Stella Nova} in the same year when he 
communicates to Sarpi his discovery of the law of falling bodies.\footnote{Letter 
on 16$^{th}$ October 1604, \textit{Opere}, vol. X, pp. 115-6, p. 115.}

\section*{The appearance of the \textit{Stella Nova}}

October 1604. The astronomers are fixing their eyes towards the 
region of the sky between the constellation of Sagittarius and 
that of the Ophiuchus or Serpentarius. They are observing quite 
a rare event, though cyclically recurrent and foreseeable: the 
celestial conjunction of three planets, Jupiter, Saturn and Mars. 
Many people are thus scrutinizing that part of the Heavens when, 
with great amazement, they suddenly see -- some say on 9$^{th}$  and 
others on 10$^{th}$ October -- a new source of light. The brightness 
of the new source of light increases during a couple of weeks 
and becomes equal to Venus. It then progressively decreases and 
finally disappears about one year and a half after its appearance. 

Different kinds of emotions shake those who observe the phenomenon: 
a mixture of astonishment and fear, of superstition and curiosity 
emerges from letters and reports of that time. People recall 
a similar appearance and disappearance of a ``stella nova'' in 
the constellation of Cassiopeia, observed in November 1572 by 
Tycho Brahe, which had raised some clamour also within the population.

What was going on? We know a lot today about these appearances. 
We can observe their remnants with our sophisticated instruments 
and we have at our disposal quite a satisfying theory of the 
stellar evolution, which enables us to catalog the appearance 
of these celestial bodies within the great class of \textit{Variable 
Stars}. It is thus sure that the phenomena observed in 1572 and 
1604 were \textit{Supernovae} (the term was introduced by Fritz Zwicky 
and Walter Baade in 1934), catastrophic events within the stellar 
evolution during which the brightness of a star suddenly increases 
so that the star becomes visible from great distances. 

In 1604, however, the knowledge was much different. The prevailing 
conception, supported by Aristotle's followers, sharply separated 
celestial phenomena and objects from terrestrial ones. 
Celestial bodies, created \textit{ab inizio} by God, were made 
of a special substance, a highly perfect quintessence that did 
not undergo through any change; their perfection was mirrored 
by the perfection of their eternal circular motions. On the contrary, 
the sublunar region, including the atmosphere and the Earth, 
was the scene of changes, of life and death, of generation and 
corruption, and it hosted bodies made of the mixture of the four 
elements (earth, water, air and fire). These bodies, according 
to the proportion of their constituting elements, had their ``natural'' place 
at a given height or distance from the centre of the Earth: if 
they were in a different position, they moved (a ``natural'' motion) 
along a straight line, to go back to their natural place. The 
downwards motion of heavy bodies (towards the Earth's surface) 
and the upwards motion of flames were explained on the basis of 
this theory.

Such a conception of the Universe, imbued with theological and 
metaphysical elements, could not fit with the appearance of new 
stars: these appearances or generations had to be linked to entities 
or bodies located not in the celestial region but in the sublunar one, they had to be meteorological phenomena, though rare 
and strange. This is why the discussion on the new star focused 
on the position of the latter. The question did not involve only 
the explanation of an event, though such a peculiar one, but 
a millenarian conception of the Heavens based on a philosophy 
of nature that had become throughout the centuries more and more 
focused on the manipulation of bibliographies, the commenting 
of books and the research of an hypothetical consistency with 
the Holy Scriptures, forgetting little by little the importance 
of direct observation. A philosophy/theology of nature which 
tried to defend itself against attacks that, from the mid 16$^{th}$ 
century, had been more and more frequent. The scientific controversy 
thus involved consolidated powers and authorities both in the 
Church and in the academic community. 

In Padua, where the \textit{nova} was observed for the first time 
on 10$^{th}$ October, the controversy was very lively and involved 
the whole town, exciting curiosity and fears among the population 
and raising careful interest among scholars. Galileo, who was 
at the time professor of mathematics and astronomy at the University 
of Padua, particularly appreciated for his teaching capacities, 
had chosen ``le teoriche dei pianeti'' as the subject of his lessons 
for the year 1604-1605. It was thus natural that his friends 
and colleagues urged him to present his opinion about the phenomenon. 
He did so on three public lessons, which were probably held from 
the end of November and the first half of December 1604. The 
curiosity was such that more than thousand persons attended the 
lessons.

Unfortunately only some notes and a few fragments of the written 
texts of these lessons still survive in the archives (assuming 
that Galileo really completely wrote down his lessons). Anyway, 
their main aim seems clear. As Galileo writes, though everybody 
was interested in knowing about ``\textit{de substantia, motu, loco 
et ratione apparitionis illius}'', he only wanted at that time 
``\textit{de motu et loco demonstrative constet}''.\footnote{\textit{Opere}, 
vol. II, p. 278.} From other sources, it is known for sure that 
Galileo intended to write down and publish his lessons. This 
is quite clear in a letter written by Alessandro Sertini to Galileo 
on 16$^{th}$ April 1605,\footnote{\textit{Opere}, vol. X, pp. 142-3, p. 143.} 
and in an unfinished letter written in January 1605 by Galileo 
to an anonymous correspondent (maybe Onofrio Castelli or, more 
probably, Girolamo Mercuriale).\footnote{\textit{Opere}, vol. X, pp. 134-5} In the latter, Galileo mentions reiterated requests to send ``copia delle tre letioni fatte da 
me in pubblico'',\footnote{``a copy of the three lessons I held in 
public'' (\textit{Ibid.}, p. 134).} and he says that the 
publication has already been repeatedly postponed and it is to 
be postponed again for a few more days, because the lessons have mainly 
dealt with the fact that the new star is much above the lunar 
orbit, while Galileo would now like to ``mutarle in discorso et 
aggiugnervi circa la sustanza et generazione''\footnote{``change them 
and add details about the substance and generation'' (\textit{Ibid.}, p. 135).} of the new star. Demonstrating that the 
star is much beyond the lunar orbit, Galileo writes, is quite ``facile, 
manifesta e comune [...]; bisogn\`{o} che io ne trattassi in grazia 
de i giovani scolari et della moltitudine bisognosa di intendere 
le demostrazioni geometriche''.\footnote{``easy, evident and common 
[...]; it was important that I presented the question to young 
students and to people who needed to hear geometrical demonstrations'' (\textit{Ibid.}, p. 134).} But discussing the substance and generation 
of the \textit{nova} was a much different matter. Galileo, in his 
letter, doesn't explicitly present his hypothesis on the subject 
(the autograph suddenly stops right with the sentence announcing 
a short summary of his ideas), he only explains that this hypothesis 
doesn't have evident contradictions and could thus be true, but 
he needs time to confirm it with observations, waiting for ``il 
ritorno di essa stella in oriente dopo la separazione del sole, 
et di nuovo osservare con gran diligenza quali mutationi abbia 
fatto s\`{\i} nel sito come nella visibile grandezza et qualit\`{a} 
di lume [...]. Et perch\'{e} questa mia fantasia si tira dietro, 
o pi\`{u} tosto si mette avanti, grandissime conseguenze et conclusioni 
per\`{o} ho risoluto di mutar le letioni in una parte di discorso''.\footnote{``the 
coming back of this star at east after the separation of the 
Sun, and observe again with great care what changes it [the star] 
shows both as for the position and the visible dimension and 
quality of light [...]. And as this \textit{fantasia} of mine brings 
extremely important consequences and conclusions forth, I have 
decided to turn the lessons into a part of a \textit{discorso}'' (\textit{Ibid.}, p. 135).}

What was this ``fantasia'' rich in consequences Galileo was working 
on? First of all, though he did not take a definitive position 
about the nature of the \textit{nova} (as he lacked indisputable evidences), 
Galileo started supporting, in those years, several hypothesis 
about the generation of the new star -- that we will discuss later 
-- that cancelled the difference between terrestrial and celestial 
physics. At the same time, Galileo hoped he could observe -- but 
he did not succeed in this -- the relative parallax of the \textit{Stella 
nova} when the Earth was at opposite positions along its revolution 
orbit around the Sun, as he thought that the changing brightness 
of the \textit{nova} was due to different distances from the Earth. 
This would have been a definitive proof that the Copernican system 
was true, against both the Ptolemaic system and the hybrid system 
proposed by Tycho Brahe. It is important to point out that in 
those very months, Galileo was working hard to study local motions, 
also in order to answer several objections to the Copernican 
system (an example: if the Earth is moving, why do we observe 
that a body falling from a tower arrives right at the base of 
the tower and not at a certain distance from it?). We can thus 
understand Galileo's emphasis about the consequences of his ``fantasia''.

\section*{\textit{The ``Dialogo de Cecco Ronchitti''}}

In Galileo's private correspondence, there are several letters 
from friends and acquaintances who sympathize with the antiaristotelian 
ideas which surely inspired the three public lessons held by 
Galileo. The Franciscan monk Ilario Altobelli, for instance, 
writes to Galileo on 3$^{rd}$ November 1604,\footnote{\textit{Opere} 
vol. X, pp 116-7.} that ``questo nuovo 
mostro del cielo''\footnote{``this new monster of the Heavens'' (\textit{Ibid.}, p. 117).} seems to be there on purpose in order 
to ``far impazzire i Peripatetici, ch'hanno creduto sin hora tante 
bugie in quella stella nova e miracolosa del 1572, priva di moto 
e di parallasse''.\footnote{``drive crazy the Peripatetics, who have 
believed so many lies, as of now, about that new and miraculous 
1572 star, motionless and without any parallax'' (\textit{Ibid.}, p. 117).} And Altobelli insisted on this point in 
a letter to Galileo written on 25$^{th}$ November 1604,\footnote{\textit{Opere}, vol. X, pp. 118-20.} where he 
repeated that the new star was clearly located on the fixed stars 
sphere and that ``il suo sito rende possibile ogni impossibilit\`{a} 
conietturata di Aristotile, distrugendo ogni sua imaginatione''\footnote{``its location makes any impossibility presupposed by Aristotle 
possible, destroying any of his ideas'', (textit{Ibid.}, p. 118}, in spite the ``pertinacia'' 
(``obstinacy'') of ``Peripatetici, o, per dir meglio, semifilosofi'' 
(``Peripatetics or, to say it better, semiphilophers''), unable 
to confront the observation data.\footnote{\textit{Ibid.}, p. 118.} And Galileo, who carried out by himself observations 
and measures on the position and features of the new star, though 
probably not in a systematic way, acquired through this intense 
correspondence, further precious details not only on the 1604 \textit{stella 
nova} but also on the previous appearances, in particular on the 
1572 one, which he was studying by reading (and commenting) Tycho 
Brahe's works. 

The antiaristotelian spirit of the three Galilean lessons raised 
a lively discussion in the Academic world, where scientific questions 
were mingled -- as often they are - with personal, prestige and 
power questions. Cesare Cremonini in particular, authoritative 
scholar of Aristotle and holder of the first chair of Natural 
Philosophy at the University of Padua, openly criticised Galileo 
and supported the Aristotelian tradition. It is likely that Cremonini 
himself inspired, at least partially, the publication in Padua, 
at the end of January 1605, of the \textit{Discorso intorno alla 
nuova stella} by Antonio Lorenzini da Montepulciano. The core 
of Lorenzini's argumentation was the strenuous defense of the 
celestial essence perfection: the immutability and incorruptibility 
of the Heavens had to imply that the \textit{nova} was nothing else 
than a meteor located in the sublunar world. To support this 
conviction, Lorenzini mentioned Aristotle, according to whom 
the Heavens would stop moving if a new star was added in it; 
he then introduced a series of 
reflections about the fact that, as the Heavens was only made 
of a quintessence, the contrary elements necessary for corruption 
and generation could not be produced in it, and he concluded 
with the rhetoric question: in what way could the Heavens corrupt 
the Heavens to generate the Heavens? After this question, he 
proposed a long digression about parallax and questions, confused 
if not even wrong, about geometric-astronomical theorems, and 
he then presented ideas from the scholastic tradition about lunar 
spots and the Via Lattea, until a further discussion on the position 
of the \textit{nova}. There were also a couple of chapters on the 
so called ``judicial astrology'', where Lorenzini discussed the 
influence of the \textit{nova} on seasons and harvests, on public 
health and on physical and moral conditions of humanity. 

An answer to Lorenzini arrived very quickly. Six weeks after 
the publication of the \textit{Discorso intorno alla nuova stella}, 
a short booklet was published in Padua with the title \textit{Dialogo 
de Cecco di Ronchitti da Bruzene in perpuosito della Stella Nuova}. 
The marginal notes of this booklet contained precise references 
to Lorenzini's text.\footnote{\textit{Opere}, vol. II, pp. 310-34.} Written 
in Paduan dialect, the \textit{Dialogo} has two main characters, Natale 
and Matteo: the first one gives an account of the ideas of a 
Paduan ``letterato'' (Lorenzini) and the other one ribs these 
ideas by using Galilean inspired arguments presented in a simple 
way and with examples taken from everyday life. It is nowadays 
ascertained that the text was written jointly by Galileo and 
Girolamo Spinelli, a young Benedictine monk of Galileo's circle. 
This circle included intellectuals and churchmen -- like the canon 
Antonio Querengo, to whom the \textit{Dialogo} is dedicated -- all 
interested not only in the new developments of science but also 
in the Paduan dialect and his great mentor, Ruzzante (alias Angelo 
Beolco). And not only the choice of the Paduan dialect is consistent 
with Beolco's ideas, but also the choice of the rough characters, 
who show how the wisdom \textit{snaturale} can prevail on the book 
based culture. 

The \textit{Dialogo}, characterised by an irony particularly manifest 
in the original dialectal version, starts with a conversation 
on the hypothetical correlation between the drought of the countryside 
and the appearance of the new star. But if it is really a star, 
says Matteo, ``as it is so far away'', it will be difficult to 
prove that it is the cause of the drought. Natale observes that 
a Paduan ``letterato'' supports in a ``librazuolo'' that the \textit{nova} 
is located in the sublunar region. Matteo then asks whether 
the author of the booklet is an expert of measures and, as he 
is told that the author ``l'\`{e} Filuorico'' (``he is a philosopher''), 
he reacts with indignation wondering ``what has his philosophy 
to do with measuring?'': the work of mathematicians is intended to carry 
out measures and they have to be asked about the position of 
the star. All right, answers Natale, the ``letterato'' also says 
that mathematicians carry out measures but they do not understand 
anything, because they have concluded from their measures that 
the star is far away and this implies an unacceptable generation 
and corruption of the Heavens. But this should not matter to 
mathematicians, answers Matteo upset, because they concern themselves 
with measuring and not with the essence of things or the substance 
of what they measure: ``even if the star was made of polenta, 
they could nevertheless observe it''. The readers of the \textit{Dialogo} 
are thus warned: the controversy on the Stella does not concern 
the simple field of astronomical observation but it involves 
the core of philosophical tradition consolidated beliefs. And 
these beliefs are to be criticised and ribbed in the following 
pages of the booklet.

Here is the argumentation proposed. Of course, Matteo argues, 
it is not possible for the moment to prove that the new star 
is really a star like all the others, but at the same time one 
can propose a series of conjectures. For instance, as it is not 
possible that ``all the stars in the Heavens could be seen'' (a 
recall to Giordano Bruno's idea), some of them could have merged 
to give birth to a new visible star, or the \textit{nova} could have 
been formed in the air and it could then have raised in the Heavens. 
As a matter of fact, though this star seems peculiar because 
of its sudden appearance, which suggests a forthcoming disappearance, 
who could support that the stars are not, like the Earth, slowly 
changing, with apparently unperceivable changes? All such arguments 
are based on the unity of the physics of the Universe, without 
any distinction between Earth and Heavens. Natale tries to answer 
to these reflections mentioning once more the ``librazuolo'', 
which says that according to Aristotle, the Heavens could not 
move any longer if a star was added. But in fact, as Matteo points 
out, this would not be such a big problem, because there are 
many people ``ed anco di buoni'' (``and good ones'') who believe 
that the Heavens does not move at all. This evident reference 
to Copernicans is explicitly written down in a marginal note in 
the Paduan edition of the \textit{Dialogo}. 

This was more than enough to drive Galileo to publish the \textit{Dialogo} under 
a pseudonym. Such a practice was common at that time, but here 
the issues were particularly delicate and they had already started 
shaking the consolidated powers within and outside the University.

\section*{The Copernican system at work}

The content of the \textit{Dialogo de Cecco di Ronchitti} and the studies 
on local motions enable Galileo to seize all the opportunities 
offered by the new instrument, the telescope. The trust in the 
observation without prejudices and the abandoning of the Aristotelian-Ptolemaic 
system are the base of his fundamental discoveries. At the same 
time, his growing trust in the Copernican system enables him 
to obtain in a natural way some of the consequences of these 
observations. In particular, as we can read at the end of the \textit{Sidereus 
Nuncius,} the discovery of Jupiter's moons is a demonstration of 
the inadequacy of those who, though accepting at first the Copernican system, become anti Copernican because they do not accept the idea that the Moon revolves around the Earth while both revolve around the Sun in one year. As a matter of fact, we now see that 
four moons are revolving around Jupiter, and all these celestial 
bodies together revolve around the Sun in twelve years. We still 
do not know how this can happen, but it happens, Galielo says.\footnote{\textit{Opere}, 
vol. III, pp. 51-96, p. 95.}

Fourteen years later, in the letter to Francesco Ingoli of 1624,\footnote{\textit{Opere}, 
vol. VI, pp. 509-561.} with his famous metaphor of the ship,\footnote{\textit{Ibid.}, pp. 
547-9.} Galileo will provide ``physics arguments''\footnote{\textit{Ibid.}, p. 
534.} to support the impossibility to prove another of the paradoxes 
against the Copernican system: if the Earth moves, how can a 
stone fall perpendicularly to the Earth's surface? Staying on 
the Earth we cannot decide if the Earth (ship) is motionless or 
in motion. The physics of local motions can help in understanding 
questions related to celestial motions. A step forward in the 
construction of a physics for the Copernican system. 

But let's go back to astronomical questions. In the ``Postscriptum'' 
of the tables on the \textit{Costitutiones of the Medicee} added to 
the \textit{Istoria e dimostrazioni intorno alle macchie solari},\footnote{\textit{Opere}, 
vol. V, pp. 247-9, p. 248.} Galileo concludes that, in order to 
explain the observed variations of the length of the eclipses 
of Jupiter's moons, it is necessary to take into account the 
fact that the shadow cone of the planet also depends on the annual 
revolution motion of the Earth [besides the dependence on the ``diverse 
latitudini di Giove'' (``different latitudes of Jupiter'') and ``dall'essere 
il pianeta che si eclissa de i pi\`{u} vicini o de' pi\`{u} lontani 
da esso Giove'' (``on the fact that the planet that is eclipsed 
may be one of the closest or most distant from Jupiter'')]. Once 
again the idea of a proof of the Copernican system. The same 
idea that will bring Galileo to hypothesise, in the same year, 
that the changing form of Saturn (sometimes with two satellites 
very close to the two opposite sides of the planet, sometimes 
alone) could depend on the relative position of the planet with 
regard to its source of illumination (the Sun) and to the observer 
(the Earth in its revolution motion).\footnote{Galileo's hypothesis 
emerges from a letter written by Agliuchi to Galileo on 13$^{th}$ 
July 1613 in answer to a letter of Galileo now lost (\textit{Opere}, 
vol. XI, pp. 532-5, p. 532).}

The adherence to observed and experimental facts and the research 
of their explanation within the most advanced scientific knowledge 
make Galileo a modern scientist. A modernity that we can see 
also in his contrariety to use Pythagoric or Platonic arguments 
so current at that time (much diffused, only to mention another 
great scientist of that time, in Kepler's work). To those who 
tried to explain, with a-priori arguments, why the moons around 
Jupiter were right four and, on the basis of these arguments, 
proposed the existence of other moons around Jupiter or around 
other planets,\footnote{See for instance the letter of Altobelli to 
Galileo on 17$^{th}$ April 1610 (\textit{Opere}, vol. X, pp. 317-8, p. 
317), and the \textit{Dissertatio} of Kepler (\textit{Opere}, vol III, 
pp. 100-25).} Galileo answers indirectly in the letter to Dini 
written on 21$^{th}$ may 1611,\footnote{\textit{Opere}, vol XI, pp. 105-16, 
p. 115.} reaffirming his adherence to facts: I have observed 
four (moons) around Jupiter and two moons around Saturn, ``non 
posso negare n\'{e} affermare cosa alcuna'' (``I cannot deny or 
affirm anything'') about whether others exist. 

With good cause, many of his contemporaries greeted Galileo as 
a new Columbus or a new Amerigo Vespucci.\footnote{Galileo is compared 
to Columbus by Orazio dal Monte (letter to Galileo on 16$^{th}$ 
June 1610, \textit{Opere}, vol. X, pp. 371-2, p. 372) and by Kepler 
in his \textit{Dissertatio} (\textit{Opere}, vol. III, p. 119), and to Amerigo 
Vespucci by Ottavio Pisani (two letters to Galileo, the first 
on 15$^{th}$ September 1613, \textit{Opere} vol. XI, p. 564-5, p. 564, 
and the second on 18$^{th}$ December 1613, \textit{Ibid.} p. 608-9, p. 
608).} A similar acknowledgement was to be addressed in 1904, 
about three hundred years after the discovery of Jupiter's moons, 
to J.J. Thomson, the scientist who discovered the electron, the 
first elementary particle.\footnote{P. Langevin, ``The Relations of Physics of Electrons to other Branches of Science'', in K. R. Sopka (ed.), \textit{Physics for a new Century. Paper Presented at the 1904 St. Louis Congress}, \textit{The History of Modern Physics 1880-1950, vol. 5, American Institute of Physics, 1986, pp. 195-230, p. 195.}}

\end{document}